\documentclass{vldb}

\usepackage[T1]{fontenc}
\usepackage[stretch=10]{microtype}
\usepackage{graphicx}
\graphicspath{{./}}
\usepackage{hyperref}
\usepackage{mathtools}
\usepackage{booktabs}
\usepackage{listings}
\usepackage{inconsolata}
\usepackage{xspace}
\usepackage{caption}
\usepackage{subcaption}
\usepackage{balance}
\usepackage{sparklines}
\definecolor{sparkrectanglecolor}{HTML}{B9F6CA}
\definecolor{sparklinecolor}{HTML}{9E9E9E}
\definecolor{sparkspikecolor}{HTML}{F44336}
\newcommand{\pr}{PRE\-STO\xspace}
\newcommand{\rs}{RPT\xspace}
\newcommand{\rss}{RPTs\xspace}

\newdef{definition}{Definition}
\newtheorem{example}{Example}

\begin{document}


%
%
%

%

\title{\pr: Probabilistic Cardinality Estimation for RDF Queries Based on Subgraph Overlapping}
%
%
%
%
%

\numberofauthors{3} 
%
\author{
%
%
\alignauthor Xin Wang\\
       \affaddr{School of Electronics and Computer Science}\\
       \affaddr{University of Southampton}\\
       \affaddr{Southampton, UK}\\
       \email{xwang@soton.ac.uk}
\alignauthor Eugene Siow\\
       \affaddr{School of Electronics and Computer Science}\\
       \affaddr{University of Southampton}\\
       \affaddr{Southampton, UK}\\
       \email{Eugene.Siow@soton.ac.uk}
\and
\alignauthor Aastha Madaan\\
       \affaddr{School of Electronics and Computer Science}\\
       \affaddr{University of Southampton}\\
       \affaddr{Southampton, UK}\\
       \email{madaan.aastha@gmail.com}
\alignauthor Thanassis Tiropanis\\
       \affaddr{School of Electronics and Computer Science}\\
       \affaddr{University of Southampton}\\
       \affaddr{Southampton, UK}\\
       \email{tt2@soton.ac.uk}
}

\additionalauthors{}
\date{}
\maketitle
\begin{abstract}
In query optimisation accurate cardinality estimation is essential for finding optimal query plans. It is especially challenging for RDF due to the lack of explicit schema and the excessive occurrence of joins in RDF queries. Existing approaches typically collect statistics based on the counts of triples and estimate the cardinality of a query as the product of its join components, where errors can accumulate even when the estimation of each component is accurate. As opposed to existing methods, we propose PRESTO, a cardinality estimation method that is based on the counts of subgraphs instead of triples and uses a probabilistic method to estimate cardinalities of RDF queries as a whole. PRESTO avoids some major issues of existing approaches and is able to accurately estimate arbitrary queries under a bound memory constraint. We evaluate PRESTO with YAGO and show that PRESTO is more accurate for both simple and complex queries.
\end{abstract}
%
%


%
%


\section{Introduction}

The Resource Description Framework (RDF)~\cite{Klyne2004} is a standard model for representing information on the Web. RDF data is a set of subject-predicate-object triples that together form a directed, labelled graph, where predicates are the edges and subjects and objects are the vertices. RDF is general and flexible. It allows both structured and semi-structured data to be mixed and shared across applications, however, it also leads to challenges when querying RDF data.

Structures in an RDF graph can be queried by specifying a conjunction of triple patterns (i.e., triples containing variables) in a query language like SPARQL~\cite{Harris2013}. Triples matched by a subset of triple patterns in a query give intermediate results, and intermediate results are joined until all triple patterns are taken into account. RDF queries typically have multiple joins and accurate cardinality estimates are critical for determining the optimal join order.


Many existing cardinality estimation approaches have two steps: 1) decomposing a query into components (e.g., triple patterns~\cite{Stocker2008,Neumann2009a,Harth2010c} or star patterns~\cite{HuangHaiLiu,Neumann2011}) and estimating cardinalities of those components; 2) estimating the cardinality of the whole query as the product of those estimations and selectivities of join conditions. There are three major issues with above approaches. First, component cardinalities and join selectivities are usually estimated based on assumptions of independence among predicates to reduce computation and storage cost, however, these assumptions do not always hold. Second, the cardinality of a query is estimated by combining estimations of components. As a result, even when the estimation of each component is accurate, errors tend to accumulate quickly for complex queries. The third issue rises from the dependency on pre-computed statistics to estimate cardinalities of components. Storage of these statistics usually increases as the size of RDF data goes up, which can be an issue when dealing with large RDF graphs.

In this paper we propose a method named \pr that is able to accurately estimate cardinalities of arbitrary complex queries without the three aforementioned issues. Considering a SPARQL query with only bound predicates, these predicates (and the way they are connected) act as the skeleton of the query and match subgraphs in a RDF graph. Any of the subgraphs that contains all the bound subjects and objects of the query is a valid result. The cardinality of the subgraphs identified by the bound vertices and the predicate skeleton is the cardinality of the query. It is prohibitive to compute in advance and store the cardinalities of all valid queries on an RDF graph, since the number of valid queries grows faster than exponentially to the number of distinct edges in the RDF graph\footnote{For $n$ distinct edges in an RDF graph, there are $C(n,k)=\binom n k$ distinct queries containing $k$ triple patterns. Each position (subject and object, assuming predicate is bound) in a triple pattern can either be bound or unbound, which gives $4^{C(n,k)}$ configurations. The total number of valid queries on the RDF graph is $\sum_{k=1}^n 4^{C(n,k)}$.}. However, the number of combinations of a single bound vertex and a predicate skeleton (i.e., queries with bound predicates and exactly one bound vertex) is much less. We refer to such a pattern as a Rooted Predicate Tree (\rs). It is realistic to store the cardinalities of the set of mostly used \rs with a Least Frequently Used (LFU) cache (least frequently used items are purged when cache is full), and to efficiently calculate cardinalities of other \rs based on cached cardinalities at runtime. In addition, a probabilistic model is proposed to calculate the most likely cardinality of a query based on the cardinality of each \rs of the query.  

In summary, \pr has the following advantages:

\begin{enumerate}
    \item Predicate dependence is captured in the cardinalities of \rs, thus no assumptions are required.
    \item Cardinality estimation is based on \rs, which can cover from a single triple pattern to a whole query. Therefore, estimation error is irrelevant to the complexity of the query (no error accumulation).
    \item Statistics storage is bound by using a combination of LFU caching and efficient runtime calculations.
\end{enumerate}

We evaluate \pr with YAGO~\cite{Suchanek2007} and show that: 1) the running time of \pr is sufficiently short even for complex queries; 2) estimations are accurate regardless of the complexity of queries.

The rest of this paper is organised as follows: Section~\ref{related} summarises existing approaches on cardinality estimation and analyses their pros and cons; Section~\ref{card} describes \pr in detail including a probabilistic model and an efficient method to collect cardinalities of \rss; the implementation of \pr is described in Section~\ref{impl}. We evaluate \pr from two perspectives, running time and accuracy, in Section~\ref{runt} and Section~\ref{accr} respectively. We conclude our work in Section~\ref{conl}. 

\section{Related Work}\label{related}


Query cardinality estimation has been intensively studied in the context of relational databases~\cite{Connell1984,Poosala1997,Ioannidis2003}. While many of the estimation techniques can be applied to RDF, they do not fully acknowledge the graph nature, heterogeneity and lack of explicit schema of RDF.

Histogram-based approaches, such as the Jena ARQ optimiser~\cite{Stocker2008}, the QTree~\cite{Harth2010c} (which combines histograms and R-Trees~\cite{AntoninGuttman1984}) and RDF-3X~\cite{Neumann2009a}, usually suffer from the correlation among predicates. Partitioning triples on joint attributes (e.g., triples matched by multiple correlated predicates) can improve estimation accuracy but the size of such histograms grows at a prohibitive speed due to combinatorial expansion.

Characteristic sets~\cite{Neumann2011} and a Bayesian-network-based approach~\cite{HuangHaiLiu} build statistics for star patterns (i.e., triple patterns sharing the same subject) instead of triple patterns. While they can provide good estimations for each star pattern, errors tends to accumulate when multiple star patterns are involved. A frequent subgraph mining approach~\cite{Maduko2008} counts potentially arbitrary subgraph patterns using the gSpan~\cite{Yan2002} algorithm, however, in practice the size of the generated statistics tends to be large due to the diversity of subgraph patterns in real-world data.

\pr differs from the existing approaches in a way that it stores only the most used subgraph patterns, based on which accurate statistics of complex graph patterns are computed at runtime. While others compromise estimation accuracy to keep statistics small, \pr provides the option to balance between running time and statistics storage while maintaining estimation accuracy for arbitrary queries.

\section{Cardinality Estimation}\label{card}

\begin{figure}[htb!]
  \centering
  \includegraphics[width=0.9\columnwidth]{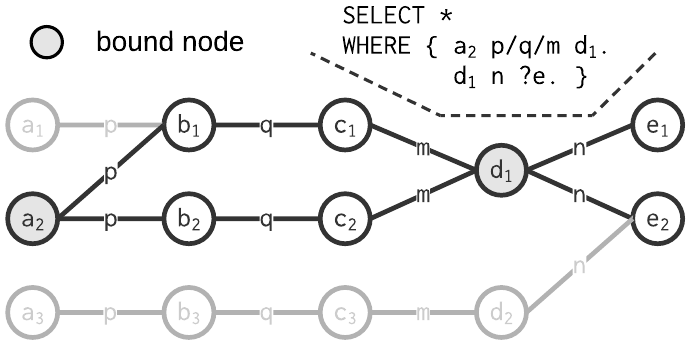}
  \caption{An RDF graph and a query matching paths passing two nodes, $a_2$ and $d_1$. Matched paths are highlighted in black.}
  \label{fig:rdf}
\end{figure}

Given the RDF graph and the query shown in Figure~\ref{fig:rdf}, the cardinality of the query equals to the number of distinct paths from $a_x$ to $e_y$ that pass both $a_2$ and $d_1$. These paths are the overlap of paths passing $a_2$ and $d_1$ respectively. Estimating the number of overlapping paths (thus the cardinality of the query) is regarded as a combinatorial problem that is to arrange the paths passing each bound node in the query (6 paths for $a_2$ and 18 for $d_1$) within the paths matched by the \rs of the query (19 paths from $a_x$ to $e_y$). The distribution of the cardinality of the query is calculated by solving the combinatorial problem under the assumption that the bound nodes (not the predicates) in the query are mutually independent (we will examine this assumption in detail in a subsequent section).

To demonstrate the cardinality distribution calculation, we firstly examine a special case that we refer to as linear graphs, and then generalise the calculation to acyclic graphs.

\subsection{Cardinality Estimation on Linear Graphs}\label{linear}

\begin{figure}[htb!]
  \centering
  \includegraphics[width=0.8\columnwidth]{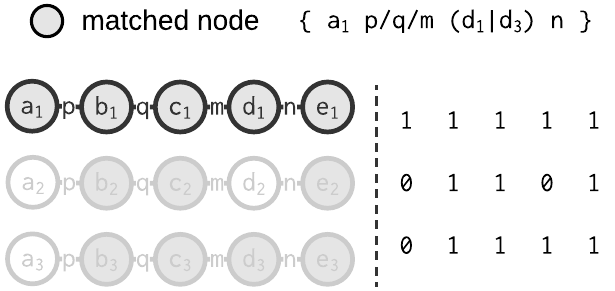}
  \caption{The linear graph of a predicate path with bound nodes and its corresponding binary matrix.}
  \label{fig:bitmap}
\end{figure}

A \emph{predicate path} is a sequence of consecutive predicates (or the inverses of them) that gives a possible route between two nodes\footnote{This definition shares a lot with a property (predicate) path described in SPARQL~1.1 Query Language~\cite{Harris2013}, but is different in the way that it focuses on consecutive predicates and does not take alternative paths into account.}. A predicate path selects components from an RDF graph (referred to as the underlining graph in the rest of this paper) which are graph paths~\cite{Weisstein}. We refer to those graph paths as a \emph{linear graph} if they are mutually disconnected, as illustrated in Figure~\ref{fig:bitmap}. 

Nodes on a predicate path can be either unbound or bound to values. We say a node in the linear graph is \emph{matched} if its value can be assigned to the corresponding node on the predicate path. We map nodes in the linear graph to a binary matrix that a node is mapped to 1 if it is matched, 0 otherwise. A predicate path without bound nodes has a binary matrix whose elements are all 1's. Figure~\ref{fig:bitmap} gives an example where the first node on the predicate path binds to $a_1$, the fourth binds to $d_1$ and $d_3$, and all other nodes are unbound. In the linear graph the path from $a_1$ to $e_1$ corresponds to an all-1 row in the matrix, and gives a valid result to the query whose body is the predicate path with bound nodes. 

It is usually prohibitive to know whether values of different variables are in the same row in the matrix. Instead we focus on the question that given the number of 1's in each column and the total number of rows, what the probability is of having $r$ all-1 rows. Unless known otherwise we assume that 1's in a column have equal chance to appear in any row. This assumption effectively changes the binary matrix of a linear graph into a \emph{Bernoulli matrix} (whose entries are iid random variables taking values from $\{0,1\}$ with probability $1/2$ each) with constraints on the sum of each column. We denote this probability as the \emph{cardinality probability} of a query. In its formal definition below we use the \emph{Iverson's bracket}~\cite[p.24]{Graham1989} notation, $[P]$, that is the $\{0,1\}$-valued function that indicates the truth of a Boolean proposition $P$, i.e.,
\[
[P]:=
    \begin{cases}
      1 & \text{if P is true,} \\
      0 & \text{otherwise.}
    \end{cases}
\]

\begin{definition}
Given a ${m \times n}$ Bernoulli matrix 
\[
 \begin{matrix}
  B_{1,1} & \dots  & B_{1,n} \\
  \vdots  & \ddots & \vdots \\
  B_{m,1} & \dots  & B_{m,n}
 \end{matrix}
\]
derived from a predicate path with bound nodes, the cardinality probability is the conditional probability \[P(T|C_1,\dots ,C_n)\]
where $T=\sum_i [R_i=n]$ is the number of all-1 rows, $R_i=\sum_j B_{i,j}$ is the sum over the $i$th row, and $C_j=\sum_i B_{i,j}$ is the sum over the $j$th column. Capital letters are ambiguously used as both random variables and their values.
\end{definition}

It is straightforward to observe the following properties:

\begin{enumerate}
\item $T$ is not affected by all-1 columns (i.e., $C_i=m$).
\item $0 \leq C_j \leq m$ and $T \leq \min(C_1,\dots,C_n)$. 
\item $T \geq C_1+C_2-m$ in a $m \times 2$ matrix.
\end{enumerate}

Property 1 states that the number of all-1 rows is only determined by the positions of the values of bound query nodes (e.g., the positions of $a_1$ in the first column and $d_1$ and $d_3$ in the fourth column). Thus the matrix in Figure~\ref{fig:bitmap} is effectively a $3 \times 2$ matrix.

\subsubsection{Cardinality Probability on 2-Column Matrices}

For a $m \times 2$ matrix the cardinality probability is expanded using Bayes' theorem as 
\[P(T|C_1,C_2)=\frac{P(C_1,C_2|T)P(T)}{P(C_1,C_2)}.\]
The probability of a column that sums to $C_j$ is \[P(C_j)=2^{-m}\binom{m}{C_j},\]
and it leads to
\begin{equation}
\label{eq:joint}
P(C_1,C_2)=P(C_1)P(C_2)=2^{-2m}\binom{m}{C_1}\binom{m}{C_2}
\end{equation}
since $C_1$ and $C_2$ are independent. Similarly the probability that a row sums to $n$ is \[P([R_i=n])=2^{-n},\] 
and it leads to
\begin{align}
P(T) & =P([R_i=2])^TP([R_i\neq 2])^{m-T}\binom{m}{T} \nonumber\\
     & =4^{-T}\left (\frac{3}{4}\right )^{m-T}\binom{m}{T}.
\label{eq:pt}
\end{align}

To calculate $P(C_1,C_2|T)$ takes more efforts than the other factors. Having two columns that sum to $C_1$ and $C_2$ respectively conditioned on forming $T$ all-1 rows in a $m$-row matrix is the same as having 2 columns that sum to $C_1-T$ and $C_2-T$ respectively conditioned on forming $0$ all-1 rows in a $(m-T)$-row matrix, i.e.,
\[ P_m(C_1,C_2|T)=P_{m'}(C'_1,C'_2|0) \]
where the subscript of a probability indicates the number of rows in the corresponding matrix, and $x'=x-T$. Applying Bayes' theorem again to the right half of the equation above leads to three factors: $P_{m'}(C'_1,C'_2)=4^{-m'}\binom{m'}{C_1'}\binom{m'}{C_2'}$ (see (\ref{eq:joint})), $P_{m'}(0)=4^{-m'}3^{m'-T'}$ (see (\ref{eq:pt})) and $P_{m'}(0|C_1',C_2')=\binom{m'-C_1'}{C_2'}/\binom{m'}{C_2'}$. The last factor is the ratio where the dividend is the number of combinations forming no all-1 rows\footnote{That is to have $C_1'$ 1's in the first column and then $C_2'$ 1's in the second column only in "empty" rows.}, i.e., $\binom{m'}{C_1'}\binom{m'-C_1'}{C_2'}$, and the divisor is the number of combinations satisfying the constraints on column sums, i.e., $\binom{m'}{C_1'}\binom{m'}{C_2'}$. The calculations above give us

\begin{equation}\label{eq:inversep}
P_m(C_1,C_2|T)=3^{T-m}\binom{m-T}{C_1-T}\binom{m-C_1}{C_2-T}.
\end{equation}
Combining (\ref{eq:joint})(\ref{eq:pt}) and (\ref{eq:inversep}) gives

\begin{equation}\label{eq:2clm}
P_m(T|C_1,C_2)=\frac{\binom{C_1}{T}\binom{m-C_1}{C_2-T}}{\binom{m}{C_2}}.
\end{equation}

Equation~(\ref{eq:2clm}) is symmetric for $C_1$ and $C_2$, i.e., exchanging $C_1$ and $C_2$ gives the same result. Intuitively it is because the order of the two columns does not affect the number of all-1 rows.

\subsubsection{Cardinality Probability on n-Column Matrices}

An n-column matrix ($n\geq3$) is treated as a series of 2-column matrices, i.e.,
\begin{equation}
\label{eq:nclm}
P(T|C_1,\dots ,C_n)=\sum_{\tilde{I}} P(\tilde{I}|C_1,C_2)P(T|\tilde{I},C_3,\dots,C_n)
\end{equation}
where $\tilde{I}\in[\max(0,C_1+C_2-m), \min(C_1,C_2)]$ is the number of intermediate all-1 rows produced by the first two columns. The second factor, $P(T|\tilde{I},C_3,\dots,C_n)$, which is the probability of forming $T$ all-q rows conditioned on that $n-1$ columns sum to $\tilde{I},C_3,\dots,C_n$ respectively, can be further expanded by applying (\ref{eq:nclm}) until there are two columns left. It corresponds to a series of triple pattern joins to produce query results.

\subsection{General Cardinality Estimation}

\begin{figure}[hbt!]
  \centering
  \includegraphics[width=\columnwidth]{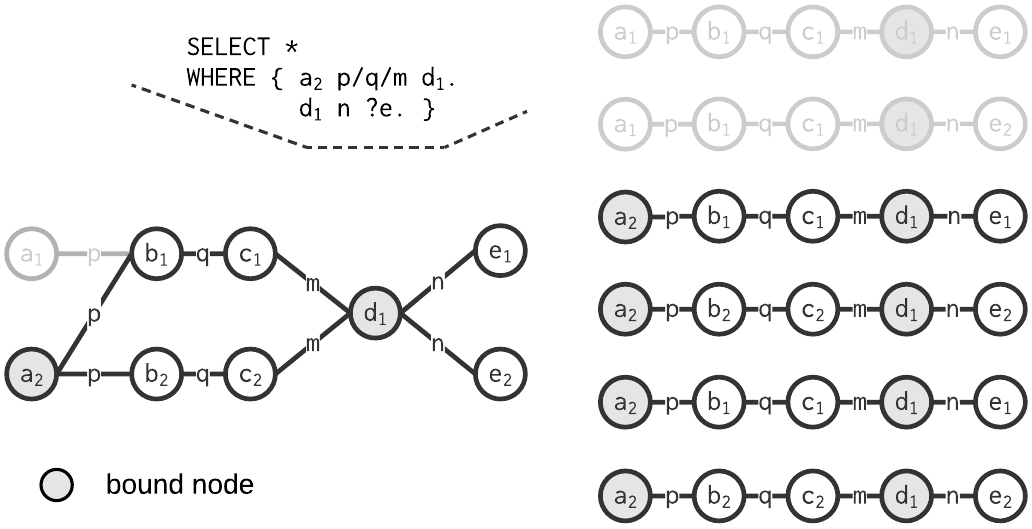}
  \caption{An acyclic RDF graph (left) mapped to a linear graph (right). Nodes that appear on more than one path are duplicated. Highlighted paths of the linear graph correspond to valid results of the query in the figure.}
  \label{fig:aclgraph}
\end{figure}

If the graph paths selected by a predicate path are not mutually disconnected (cf.\ Section~\ref{linear}), we convert those paths to a linear graph by duplicating nodes where different paths join to split connected paths, as demonstrated in Figure~\ref{fig:aclgraph}. After the conversion a bound node on the predicate path can appear on multiple paths in the linear graph. To apply (\ref{eq:nclm}) to the linear graph we need to know the number of occurrences of each bound node, e.g., $C_1$ should be 4 in this case since $a_2$ occurs four times.

A query can match subgraphs that are trees (that are edge-preserving isomorphic) instead of paths, and it is not straightforward to unambiguously map trees to a linear graph. In this case we discard linear graphs, despite being a good visual aid, and generalise the interpretation of (\ref{eq:2clm}). In general the Bernoulli matrix represents the results of a query more than the subgraphs matched by the query. The number of matching subgraphs is the number of rows ($m$), and the numbers of occurrences of values in the matching subgraphs are the constraints on column sums ($C_i$). These numbers are calculated by means of \rs.

\subsection{\rs Cardinality Calculation}

\begin{figure}[hbt!]
  \centering
  \includegraphics[width=\columnwidth]{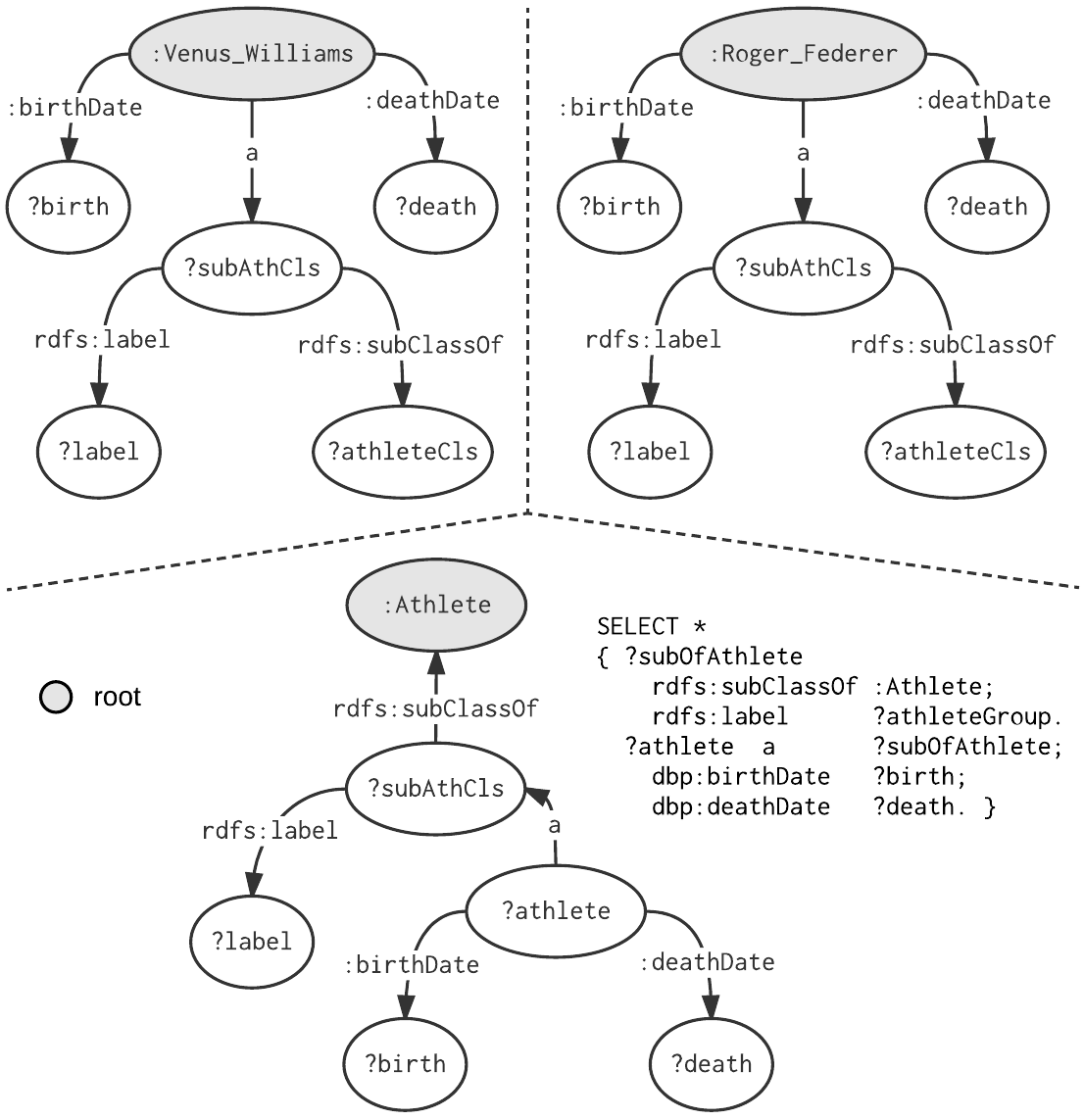}
  \caption{Three possible \rss of the same query. The top left and top right \rss have the same root node but bound to different values. The bottom \rs roots at a different node.}
  \label{fig:rpt}
\end{figure}

An \rs of a query is a rooted tree~\cite{Weissteina} whose edges are all the predicates of the query, and only the root is bound to a value. A query has many \rs that are uniquely identified by the positions and values of the root nodes. Figure~\ref{fig:rpt} gives an example of three possible \rss of a query.

Each \rs matches trees in an RDF graph whose isomorphisms preserve the \rs's root value and edges. These trees form a set that we refer to as the \emph{\rs tree set} of the root value. Given a query, the intersection (overlapping) of the \rs tree sets of all concrete values gives the query results. The cardinality of an \rs tree set gives the number of occurrences of the root value in the Bernoulli matrix and thus the constraints on column sums. Summing over all possible values, drawn from the underlining RDF, of an \rs root gives the number of rows of the Bernoulli matrix ($m$).

An na\"ive way to calculate \rs cardinalities is to modify the original query to have the root as the only concrete node, and executing the modified query. Apparently it would be time consuming and defy our purpose of estimating cardinalities for complex queries. Here we propose a method that breaks \rs cardinality calculations into small reusable parts that are stored in cache. Later calculations can reuse partial results from the cache to save (a significant amount of) time.  

Noting the recursive structure of an \rs, its cardinality is the aggregation of the cardinalities of sub-\rss (subtrees) rooted at the neighbours of its root. The aggregation follows the rules below (thinking of how query results are constructed):

\begin{itemize}
    \item \textbf{Base case}: A single-node \rs has cardinality 1. 
    \item \textbf{Addition rule}: Cardinalities of sub-\rss descending from the same predicate are summed.
    \item \textbf{Multiplication rule}: Cardinalities of sub-\rss descending from different predicates are multiplied.
\end{itemize} 
Denoting an \rs as an ordered pair $(r,T)$ where $r$ is the root and $T$ is the tree of predicates, and the cardinality of the \rs as $|r,T|$, the above rules are summarised in the following equation
\begin{equation}\label{eq:card}
|r,T|=
    \begin{dcases*}
        ~~~~1 & if $T=\emptyset$, \\
        \prod_{e \in E(r,T)}~\sum_{v \in V(r,e)} |v,T \sim e| & otherwise.
    \end{dcases*}
\end{equation}
where $E(r,T)$ gives all edges in $T$ that are adjacent to $r$, $V(r,e)$ gives all the nodes in the underlining RDF graph that are adjacent to $r$ following $e$, and $T \sim e$ is the subtree of $T$ following $e$.

\begin{example}[\rs Cardinality]
The cardinality of $(a_2,p/q/m/n)$, in Figure~\ref{fig:aclgraph}, is the sum of paths descending from the predicate $p$, i.e.,
\[\underbrace{|a_2,p/q/m/n|}_{4}=\underbrace{|b_1,q/m/n|}_{2}+\underbrace{|b_2,q/m/n|}_{2};\]
the cardinality of $(d_1, p/q/m~d_1~n)$ is given by the product of two sums, i.e.,
\[\underbrace{|d_1, p/q/m~d_1~n|}_{6}=(\underbrace{|c_1, q/p|}_{2}+\underbrace{|c_2,q/p|}_{1})(\underbrace{|e_1,\emptyset|}_{1}+\underbrace{|e_2,\emptyset|}_{1}).\]
\end{example}

Equation (\ref{eq:card}) translates straight to an recursive algorithm (omitted here) once all operations are translated to pseudo-code. Memoisation~\cite{Michie1968} is used to reduce the complexity of the calculation, i.e., \rs cardinalities calculated in the past are cached, and (\ref{eq:card}) goes to the next level of recursion only when $|r,T|$ is not found in the cache. It would be space consuming to store all historical cardinalities and unnecessary since not all \rss would reoccur with the same probability. We store only the mostly used \rss in an LFU cache in this case as a way to balance between the time and space complexity of (\ref{eq:card}).

\section{Implementation}\label{impl}

We implement our methods as \pr with Mathematica\footnote{\url{https://www.wolfram.com/mathematica}} and Jena\footnote{\url{https://jena.apache.org}} (a Java based semantic web platform). The main components of \pr are straight translations of (\ref{eq:nclm}) into Mathematica and (\ref{eq:card}) into Java respectively, however, clarifications on relatively minor components would benefit interested users to reproduce \pr. In addition, we provide the source code at \url{https://github.com/xgfd/Presto}.

\subsection{Representing \rss}

\rs is implemented as a pair, the root and the connected edges (i.e., $(r,T)$), the same way as we write it in this paper. The root is implemented with the \emph{Node} class in Jena. 

The data structure of $T$ should efficiently support the following operations: 1) a method that gives the subtree of a $T$ descending from an edge (i.e., $T\sim e$); 2) a method that gives the equality of two \rss that overrides the \emph{equals()} method for retrieving memorised cardinalities; 3) a hash function that overrides the \emph{hashCode()} method for retrieving memorised cardinalities. With these requirements in mind, $T$ is implemented as a collection of directed edges pointing to other $Ts$\footnote{Path in the repository: \href{https://github.com/xgfd/Presto/blob/master/src/ELT.java}{src/ELT.java}}. This gives a fast $O(1)$ implementation of $T\sim e$. This implementation allows the reuse of the connected edges among \rss that are only different at their roots, and gives us a shortcut to compare the equality of two \rss. 

Tree equality (isomorphism) has been well studied~\cite{Campbell1991a,Carroll2002,Kim2015} and can be solved by the AHU algorithm by Aho, Hopcroft, and Ullma~\cite{Aho1974}. However, the AHU algorithm is not straightforward to understand nor to implement. We notice that given a query the connected edges of each \rs only needs to be generated once, therefore two \rss are equal i.i.f their Java references of the root and the collection of edges are the same, testable by the built-in \emph{equals()} method.

$T$ is treated as a set of linked list of edges in generating the hash code, i.e., it sums the products of sub-$Ts$ and the edges leading to them, where incoming and outgoing edges are given different (prime) weights.

A SPARQL query is transformed into an \rs by a depth-first traversal, starting from a (bound) node in the query.

\subsection{Adjacent Edges and Nodes}\label{adjnodes}

In (\ref{eq:card}) $E(r,T)$ retrieves adjacent edges of $r$ in $T$ and $V(r,e)$ gives all adjacent nodes of $r$ in the underlining RDF via $e$. The implementation of the former is straightforward which returns the collection of edges in $T$. The implementation of the later has multiple options. A fast approach is to store adjacent nodes in a hash table that each node links to all its neighbours. However this method replicates the RDF graph. In \pr we reuse the RDF store provided by Jena and implements $V(r,e)$ as a query in the form of ${r~e~?v}$ or ${?v~e~r}$ depending on the direction of $e$. Memoisation is also used in $V(r,e)$ to improve performance.

\subsection{Caching \rs Cardinalities}

The cache of \rs cardinalities is implemented based on an $O(1)$ LFU algorithm~\cite{Shah2010}. It can be configured with two parameters, the maximum size of the cache and the eviction rate, which allow us to tune the trade-off between the time and space complexity of \pr.

\section{Running Time Evaluation}\label{runt}

\pr performs a certain amount of computation at runtime due to the complexity of (\ref{eq:nclm}) and (\ref{eq:card}), and we are interested to know whether \pr is sufficiently fast for query optimisation.

The structure of RDF data and queries can affect \pr's performance in complex ways, which leads our choice to real-world data as artificial data tend to be limited in their structuredness~\cite{Duan2011}. We also struggle to find in existing benchmarks queries that cover a variety of complexity and structures. With the aim to give a relatively realistic and comprehensive evaluation we decide to use the YAGO dataset and construct our own queries.

\subsection{Evaluation Setup}\label{evalquery}
We construct templates covering star-shaped (Q1--4) and property path (Q5--7) queries, as listed in Table~\ref{tab:template}. Each template contains 2-3 parameters (marked with "\$") that are substituted with values from the YAGO dataset to generate testing queries. More specifically, parameter values are drawn independently from the subgraph that matches the predicates of the query (i.e., change all nodes in the query to variables). 

\begin{table}[!htb]
\caption{Query templates}
    \label{tab:template}
\begin{tabular}{@{}l l@{}}
    \textbf{BASE}   & \textbf{<http://mpii.de/yago/resource/>}\\
    \toprule
    \textbf{Q1} 
                    & 
                        \begin{minipage}{0.8\columnwidth}
\begin{lstlisting}
SELECT  *
{ ?s <bornIn>   $1;
     rdfs:label ?n;
     a          $2
}
\end{lstlisting}
                        \end{minipage}
    \\ 
    \midrule
    \textbf{Q2} 
                    &
                        \begin{minipage}{0.8\columnwidth}
\begin{lstlisting}
SELECT  *
{ ?s <originatesFrom> $1;
     rdfs:label       ?n;
     <hasWebsite>     $2 
}
\end{lstlisting}
                        \end{minipage}
    \\ 
    \midrule
    \textbf{Q3} 
                    & 
                        \begin{minipage}{0.8\columnwidth}
\begin{lstlisting}
SELECT  *
{ ?s <diedIn>     $1;
     rdfs:label   ?n;
     <diedOnDate> ?d;
     a            $2 
}
\end{lstlisting}
                        \end{minipage}
    \\ 
    \midrule
    \textbf{Q4} 
                    & 
                        \begin{minipage}{0.8\columnwidth}
\begin{lstlisting}
SELECT  *
{ ?s <hasOfficialLanguage> $1;
     <hasUTCOffset>        $2;
     <hasCapital>          ?c;
     <hasCurrency>         ?cur 
}
\end{lstlisting}
                        \end{minipage}
    \\ 
    \midrule
    \textbf{Q5} 
                    & 
                        \begin{minipage}{0.8\columnwidth}
\begin{lstlisting}
SELECT  *
{ $2 <produced>/
       <isOfGenre>/
         rdfs:label ?label;
     <influences>   $1 
}
\end{lstlisting}
\end{minipage}
    \\
    \midrule
    \textbf{Q6} 
                    & 
                        \begin{minipage}{0.8\columnwidth}
\begin{lstlisting}
SELECT  *
{ ?director <directed>     $1;
            <influences>   $3;
            <produced>/ 
              <isOfGenre>/ 
                rdfs:label ?label.
  $1 <hasPredecessor>/
       <hasPredecessor>/ 
         a $2 
}
\end{lstlisting}
                        \end{minipage}
    \\
    \midrule
    \textbf{Q7} 
                    & 
                        \begin{minipage}{0.8\columnwidth}
\begin{lstlisting}
SELECT  *
{ $1 <discovered>/ 
       <influences>/ 
         <discovered>/
           <influences>/
             <bornOnDate> $2
}
\end{lstlisting}
                        \end{minipage}
    \\ 
    \bottomrule
\end{tabular}
\end{table}

We generate 50 queries each from Q1--3, 25 queries from Q4, 44 queries from Q5, 9 queries from Q6 and 100 queries from Q7. The variance of the number of queries from each template is due to the different numbers of values per parameters in a template, i.e., more complex queries tend to have less values per parameter except Q7. All generated queries can be found in the \pr GitHub repository\footnote{Path in the repository: \href{https://github.com/xgfd/Presto/tree/master/test/yago_queries}{test/yago\_queries}}.

All tests are run on an iMac with a 3.2~GHz Intel Core i5 processor and 8~GB 1600~MHz DDR3 memory. The initial and maximum heap size of Java are set to 1000~MB and 3000~MB respectively (i.e., -Xms1000m -Xmx3000m).

\subsection{Results and Analysis}

This evaluation has 5 measurements: 1) the time calculating \rs cardinalities (\ref{eq:card}); 2) the time calculating cardinality probabilities (\ref{eq:nclm}); 3) the number of cache hits; 4) the number of cache misses, and 5) the ratio between hits and misses. The results are averaged for each query template and presented in Table~\ref{tab:runningtime}. In addition the cardinality and probability calculation time of individual queries are given as sparklines.

\begin{table*}[!hbt]
    \caption {\pr running time \& caching}
    \label{tab:runningtime}
    \centering
    \begin{tabular}{@{}rrrrrrrrr@{}}
    ~  & \multicolumn{5}{c}{Running Time (ms)}                                    & \multicolumn{3}{c}{Cache} \\ 
         \cmidrule(lr){2-6}                                                        \cmidrule(lr){7-9}
    ~  & \multicolumn{2}{c}{\textbf{Card. (Eq. \ref{eq:card})}} & \multicolumn{2}{c}{\textbf{Prob. (Eq. \ref{eq:nclm})}} & \textbf{Total} & \textbf{Hit} & \textbf{Miss} & \textbf{Hit/Miss} \\ 
    \toprule
    \textbf{Q1} & 1902 &
                        \begin{sparkline}{12}
                        \sparkdot 0.44 0.29 blue
                        \sparkdot 0.08 0.91 red
                        \spark 0.02 0.37 0.04 0.72 0.06 0.55 0.08 0.91 0.10 0.50 0.12 0.51 0.14 0.43 0.16 0.40 0.18 0.49 0.20 0.49 0.22 0.48 0.24 0.67 0.26 0.46 0.28 0.50 0.30 0.43 0.32 0.51 0.34 0.53 0.36 0.50 0.38 0.51 0.40 0.46 0.42 0.70 0.44 0.29 0.46 0.33 0.48 0.49 0.50 0.47 0.52 0.46 0.54 0.49 0.56 0.45 0.58 0.36 0.60 0.48 0.62 0.45 0.64 0.48 0.66 0.47 0.68 0.45 0.70 0.37 0.72 0.49 0.74 0.47 0.76 0.47 0.78 0.47 0.80 0.36 0.82 0.50 0.84 0.49 0.86 0.51 0.88 0.51 0.90 0.55 0.92 0.54 0.94 0.48 0.96 0.42 0.98 0.36 /
                        \end{sparkline}        
                            & 447 &
                                   \begin{sparkline}{12}
                                   \sparkdot 0.06 0.22 blue
                                   \sparkdot 0.54 0.91 red
                                   \spark 0.02 0.29 0.04 0.29 0.06 0.22 0.08 0.37 0.10 0.33 0.12 0.32 0.14 0.30 0.16 0.28 0.18 0.33 0.20 0.38 0.22 0.31 0.24 0.32 0.26 0.29 0.28 0.28 0.30 0.38 0.32 0.44 0.34 0.46 0.36 0.41 0.38 0.71 0.40 0.56 0.42 0.51 0.44 0.51 0.46 0.55 0.48 0.53 0.50 0.58 0.52 0.57 0.54 0.91 0.56 0.56 0.58 0.54 0.60 0.56 0.62 0.57 0.64 0.57 0.66 0.55 0.68 0.56 0.70 0.55 0.72 0.56 0.74 0.58 0.76 0.56 0.78 0.57 0.80 0.54 0.82 0.56 0.84 0.54 0.86 0.56 0.88 0.58 0.90 0.56 0.92 0.51 0.94 0.55 0.96 0.55 0.98 0.55 /
                                   \end{sparkline}        
                                    & 2349 & 150662 & 307933 & 0.49     \\
    \textbf{Q2} & 162  & 
                        \begin{sparkline}{12}
                        \sparkdot 0.71 0.10 blue
                        \sparkdot 0.10 0.91 red
                        \spark 0.01 0.91 0.03 0.36 0.05 0.18 0.07 0.22 0.09 0.62 0.11 0.69 0.13 0.11 0.15 0.14 0.17 0.19 0.19 0.22 0.21 0.26 0.23 0.57 0.25 0.17 0.27 0.15 0.29 0.19 0.31 0.23 0.33 0.26 0.35 0.51 0.37 0.12 0.39 0.23 0.41 0.19 0.43 0.29 0.45 0.32 0.47 0.68 0.49 0.13 0.51 0.17 0.53 0.20 0.55 0.24 0.57 0.28 0.59 0.54 0.61 0.20 0.63 0.23 0.65 0.39 0.67 0.22 0.69 0.60 0.71 0.10 0.73 0.19 0.75 0.18 0.77 0.21 0.79 0.25 0.81 0.68 0.83 0.16 0.85 0.20 0.87 0.18 0.89 0.27 0.91 0.25 0.93 0.60 0.95 0.11 0.97 0.15 0.99 0.18 /
                        \end{sparkline}        
                            & 406 & 
                                   \begin{sparkline}{12}
                                   \sparkdot 0.01 0.10 blue
                                   \sparkdot 0.69 0.91 red
                                   \spark 0.01 0.10 0.03 0.28 0.05 0.31 0.07 0.26 0.09 0.21 0.11 0.20 0.13 0.45 0.15 0.48 0.17 0.53 0.19 0.26 0.21 0.33 0.23 0.28 0.25 0.26 0.27 0.49 0.29 0.47 0.31 0.51 0.33 0.49 0.35 0.50 0.37 0.50 0.39 0.52 0.41 0.55 0.43 0.32 0.45 0.38 0.47 0.83 0.49 0.39 0.51 0.80 0.53 0.78 0.55 0.79 0.57 0.37 0.59 0.79 0.61 0.76 0.63 0.79 0.65 0.31 0.67 0.31 0.69 0.91 0.71 0.33 0.73 0.29 0.75 0.39 0.77 0.33 0.79 0.32 0.81 0.27 0.83 0.29 0.85 0.29 0.87 0.30 0.89 0.32 0.91 0.33 0.93 0.32 0.95 0.30 0.97 0.30 0.99 0.34 /
                                   \end{sparkline}        
                                    & 568  & 1558   & 51505  & 0.03     \\
    \textbf{Q3} & 1519 & 
                        \begin{sparkline}{12}
                        \sparkdot 0.49 0.12 blue
                        \sparkdot 0.11 0.91 red
                        \spark 0.01 0.61 0.03 0.67 0.05 0.32 0.07 0.46 0.09 0.58 0.11 0.91 0.13 0.68 0.15 0.67 0.17 0.33 0.19 0.14 0.21 0.17 0.23 0.13 0.25 0.69 0.27 0.68 0.29 0.74 0.31 0.67 0.33 0.69 0.35 0.26 0.37 0.24 0.39 0.13 0.41 0.17 0.43 0.14 0.45 0.16 0.47 0.20 0.49 0.12 0.51 0.70 0.53 0.69 0.55 0.45 0.57 0.57 0.59 0.68 0.61 0.32 0.63 0.23 0.65 0.18 0.67 0.14 0.69 0.69 0.71 0.32 0.73 0.22 0.75 0.72 0.77 0.32 0.79 0.49 0.81 0.33 0.83 0.51 0.85 0.66 0.87 0.33 0.89 0.13 0.91 0.19 0.93 0.53 0.95 0.33 0.97 0.53 0.99 0.32 /
                        \end{sparkline}        
                            & 370 &  
                                   \begin{sparkline}{12}
                                   \sparkdot 0.11 0.17 blue
                                   \sparkdot 0.89 0.91 red
                                   \spark 0.01 0.87 0.03 0.24 0.05 0.19 0.07 0.28 0.09 0.22 0.11 0.17 0.13 0.47 0.15 0.36 0.17 0.43 0.19 0.41 0.21 0.35 0.23 0.41 0.25 0.41 0.27 0.40 0.29 0.40 0.31 0.47 0.33 0.50 0.35 0.52 0.37 0.43 0.39 0.43 0.41 0.52 0.43 0.51 0.45 0.53 0.47 0.49 0.49 0.50 0.51 0.44 0.53 0.37 0.55 0.50 0.57 0.74 0.59 0.85 0.61 0.85 0.63 0.88 0.65 0.88 0.67 0.81 0.69 0.85 0.71 0.83 0.73 0.84 0.75 0.84 0.77 0.83 0.79 0.78 0.81 0.83 0.83 0.78 0.85 0.87 0.87 0.82 0.89 0.91 0.91 0.80 0.93 0.84 0.95 0.84 0.97 0.80 0.99 0.79 /
                                   \end{sparkline}        
                                    & 1889 & 86789  & 202841 & 0.43     \\
    \textbf{Q4} & 4    & 
                        \begin{sparkline}{12}
                        \sparkdot 0.74 0.30 blue
                        \sparkdot 0.02 0.91 red
                        \spark 0.02	0.91 0.06 0.51 0.10	0.40 0.14 0.30 0.18	0.30 0.22	0.51 0.26 0.30 0.30	0.30 0.34 0.30 0.38	0.30 0.42 0.30 0.46	0.30 0.50 0.30 0.54	0.30 0.58 0.30 0.62	0.51 0.66 0.40 0.70	0.40 0.74 0.30 0.78	0.40 0.82 0.40 0.86	0.40 0.90 0.40 0.94	0.40 0.98 0.51 /
                        \end{sparkline}        
                            & 116 &  
                                   \begin{sparkline}{12}
                                   \sparkdot 0.14 0.12 blue
                                   \sparkdot 0.70 0.91 red
                                   \spark 0.02 0.13 0.06 0.12 0.10 0.16 0.14 0.12 0.18 0.13 0.22 0.74 0.26 0.74 0.30 0.73 0.34 0.64 0.38 0.67 0.42 0.42 0.46 0.35 0.50 0.36 0.54	0.55 0.58 0.87 0.62	0.68 0.66 0.54 0.70	0.91 0.74 0.68 0.78 0.53 0.82 0.53 0.86 0.18 0.90	0.75 0.94 0.47 0.98 0.53 /
                                   \end{sparkline}        
                                    & 120  & 243    & 565    & 0.43     \\
    \textbf{Q5} & 60   & 
                        \begin{sparkline}{12}
                        \sparkdot 0.06 0.08 blue
                        \sparkdot 0.91 0.91 red
                        \spark 0.01 0.09 0.03 0.27 0.06 0.08 0.08 0.09 0.10 0.10 0.13 0.09 0.15 0.10 0.17 0.11 0.20 0.11 0.22 0.13 0.24 0.10 0.26 0.40 0.29 0.11 0.31 0.10 0.33 0.18 0.36 0.13 0.38 0.13 0.40 0.10 0.43 0.14 0.45 0.15 0.47 0.15 0.49 0.10 0.52 0.15 0.54 0.12 0.56 0.14 0.59 0.13 0.61 0.19 0.63 0.20 0.66 0.26 0.68 0.21 0.70 0.28 0.72 0.21 0.75 0.22 0.77 0.32 0.79 0.29 0.82 0.27 0.84 0.39 0.86 0.31 0.89 0.27 0.91 0.91 0.93 0.25 0.95 0.22 0.98 0.23 1.00 0.20 /
                        \end{sparkline}        
                            & 321 &  
                                   \begin{sparkline}{12}
                                   \sparkdot 0.15 0.06 blue
                                   \sparkdot 0.79 0.91 red
                                   \spark 0.01 0.15 0.03 0.14 0.06 0.25 0.08 0.12 0.10 0.09 0.13 0.07 0.15 0.06 0.17 0.08 0.20 0.07 0.22 0.08 0.24 0.07 0.26 0.08 0.29 0.08 0.31 0.08 0.33 0.08 0.36 0.08 0.38 0.08 0.40 0.07 0.43 0.09 0.45 0.15 0.47 0.23 0.49 0.18 0.52 0.34 0.54 0.30 0.56 0.27 0.59 0.33 0.61 0.49 0.63 0.40 0.66 0.40 0.68 0.28 0.70 0.36 0.72 0.34 0.75 0.35 0.77 0.81 0.79 0.91 0.82 0.40 0.84 0.74 0.86 0.72 0.89 0.39 0.91 0.56 0.93 0.50 0.95 0.79 0.98 0.37 1.00 0.37 /
                                   \end{sparkline}        
                                    & 381  & 6441   & 7298   & 0.88     \\
    \textbf{Q6} & 58   &  
                        \begin{sparkline}{12}
                        \sparkdot 0.37 0.16 blue
                        \sparkdot 0.25 0.91 red
                        \spark 0.12	0.63 0.25 0.91 0.37	0.16 0.49 0.71 0.62	0.19 0.74	0.72 0.86 0.65 0.98	0.29 /
                        \end{sparkline}        
                            & 78  &  
                                   \begin{sparkline}{12}
                                   \sparkdot 0.86 0.70 blue
                                   \sparkdot 0.12 0.91 red
                                   \spark 0.12 0.91 0.25 0.77 0.37 0.76 0.49 0.73 0.62 0.72 0.74 0.71 0.86 0.70 0.98 0.74 /
                                   \end{sparkline}        
                                    & 136  & 4178   & 4336   & 0.96     \\
    \textbf{Q7} & 38   &  
                        \begin{sparkline}{12}
                        \sparkdot 0.03 0.01 blue
                        \sparkdot 0.78 0.91 red
                        \spark 0.02 0.03 0.03 0.01 0.04 0.20 0.05 0.02 0.06 0.03 0.07 0.02 0.08 0.02 0.09 0.48 0.10 0.18 0.11 0.01 0.12 0.01 0.13 0.01 0.14 0.17 0.15 0.16 0.16 0.18 0.17 0.01 0.18 0.44 0.19 0.01 0.20 0.22 0.21 0.02 0.22 0.22 0.23 0.02 0.24 0.20 0.25 0.23 0.26 0.02 0.27 0.27 0.28 0.46 0.29 0.02 0.30 0.22 0.31 0.25 0.32 0.02 0.33 0.02 0.34 0.26 0.35 0.02 0.36 0.02 0.37 0.26 0.38 0.26 0.39 0.27 0.40 0.34 0.41 0.64 0.42 0.46 0.43 0.03 0.44 0.77 0.45 0.04 0.46 0.03 0.47 0.81 0.48 0.38 0.49 0.02 0.50 0.28 0.51 0.02 0.52 0.02 0.53 0.03 0.54 0.03 0.55 0.03 0.56 0.33 0.57 0.03 0.58 0.33 0.59 0.04 0.60 0.03 0.61 0.03 0.62 0.03 0.63 0.32 0.64 0.04 0.65 0.03 0.66 0.33 0.67 0.04 0.68 0.04 0.69 0.04 0.70 0.34 0.71 0.04 0.72 0.36 0.73 0.04 0.74 0.36 0.75 0.04 0.76 0.03 0.77 0.05 0.78 0.91 0.79 0.04 0.80 0.33 0.81 0.04 0.82 0.35 0.83 0.04 0.84 0.36 0.85 0.36 0.86 0.03 0.87 0.05 0.88 0.36 0.89 0.38 0.90 0.42 0.91 0.40 0.92 0.40 0.93 0.41 0.94 0.42 0.95 0.05 0.96 0.04 0.97 0.05 0.98 0.04 0.99 0.05 /
                        \end{sparkline}        
                            & 97  &  
                                   \begin{sparkline}{12}
                                   \sparkdot 0.30 0.22 blue
                                   \sparkdot 0.19 0.91 red
                                   \spark 0.02 0.26 0.03 0.26 0.04 0.27 0.05 0.42 0.06 0.35 0.07 0.35 0.08 0.27 0.09 0.28 0.10 0.26 0.11 0.27 0.12 0.27 0.13 0.26 0.14 0.27 0.15 0.48 0.16 0.25 0.17 0.24 0.18 0.28 0.19 0.91 0.20 0.23 0.21 0.25 0.22 0.24 0.23 0.24 0.24 0.22 0.25 0.23 0.26 0.26 0.27 0.28 0.28 0.33 0.29 0.25 0.30 0.22 0.31 0.22 0.32 0.67 0.33 0.25 0.34 0.24 0.35 0.25 0.36 0.26 0.37 0.27 0.38 0.22 0.39 0.22 0.40 0.22 0.41 0.24 0.42 0.22 0.43 0.32 0.44 0.26 0.45 0.31 0.46 0.65 0.47 0.30 0.48 0.22 0.49 0.25 0.50 0.23 0.51 0.65 0.52 0.26 0.53 0.25 0.54 0.28 0.55 0.25 0.56 0.28 0.57 0.25 0.58 0.23 0.59 0.68 0.60 0.25 0.61 0.25 0.62 0.62 0.63 0.22 0.64 0.25 0.65 0.66 0.66 0.25 0.67 0.68 0.68 0.24 0.69 0.68 0.70 0.25 0.71 0.69 0.72 0.25 0.73 0.66 0.74 0.25 0.75 0.66 0.76 0.51 0.77 0.27 0.78 0.32 0.79 0.24 0.80 0.22 0.81 0.23 0.82 0.22 0.83 0.24 0.84 0.26 0.85 0.22 0.86 0.64 0.87 0.25 0.88 0.23 0.89 0.22 0.90 0.25 0.91 0.22 0.92 0.22 0.93 0.22 0.94 0.22 0.95 0.24 0.96 0.24 0.97 0.66 0.98 0.25 0.99 0.24 /
                                   \end{sparkline}
                                    & 135  & 2498   & 4120   & 0.60     \\
    \bottomrule
    \end{tabular}
\end{table*}

The total running time of (\ref{eq:nclm}) and (\ref{eq:card}) ranges from 120~ms to 2349~ms: Q4--7 are under 500~ms, Q2 is around 500~ms, and Q1,3 are around 2000~ms. Over 500~ms appears to be too long for cardinality estimation, and we investigate further for possible causes and improvement.

Calculating (\ref{eq:nclm}) itself can be slow due to the latency of the communication between Java and Mathematica. This latency can be removed by implementing (\ref{eq:nclm}) in Java. It is also worth mentioning that even \pr gives a single number as the most probable cardinality (cacheable), it keeps the cardinality distribution (non-cacheable) given by (\ref{eq:nclm}). It is straightforward to cache the most probable cardinality and (\ref{eq:nclm}) is only calculated for the first time or when cardinality distribution is needed to make finer estimations.

The time complexity of (\ref{eq:card}) largely depends on the number of intermediate results. Q1--3 are simple star-shaped queries with less selective predicates---as opposed to Q4 that has more selective predicates and Q5--7 that are path queries---and generate 1--2 magnitudes more intermediate results than the rest. This is also reflected in the number of cache accesses. As described in Section~\ref{adjnodes}, the current implementation uses a slow method (by querying the underlining RDF) to retrieve adjacent nodes. Experiments show that by using a hash-map to store adjacent nodes reduces the time of (\ref{eq:card}) to under tens of milliseconds, although it does not scale well with large RDF datasets. In addition we argue that \rs cardinalities are also side products of the query execution phase, and can be used by \pr in later estimation without calculating (\ref{eq:card}). In other words, the complexity of (\ref{eq:card}) is negligible when it is amortised with query execution.

In summary we conclude that \pr can be sufficiently fast for even non-selective queries with more sophisticated implementations. 

\subsubsection{Performance Improved by Caching}

We evaluate the effectiveness of caching by comparing the performance of \pr with or without cache, as presented in Table~\ref{tab:cache}. Only the cardinality calculation time is recorded since probability calculation is not cached. When the cache is turned off cache misses are simply the number of attempts to access the cache. The ratios of results of the two situations are given in the last double-column to show the degree of performance improvement.

\begin{table}[!htb]
\centering
\caption{Performance improved by caching}
\label{tab:cache}
\begin{tabular}{@{}rrrrrrr@{}}
            & \multicolumn{2}{c}{Without cache} & \multicolumn{2}{c}{With cache} & \multicolumn{2}{c}{Ratio}      \\
            \cmidrule(lr){2-3} \cmidrule(lr){4-5} \cmidrule(lr){6-7}
            & \textbf{Card.}   & \textbf{Miss}  & \textbf{Card.} & \textbf{Miss} & \textbf{Card.} & \textbf{Miss} \\
            \toprule
\textbf{Q1} & 4325             & 982622         & 1902           & 307933        & 2.27           & 3.19          \\
\textbf{Q2} & 165              & 53090          & 162            & 51505         & 1.02           & 1.03          \\
\textbf{Q3} & 3963             & 533272         & 1519           & 202841        & 2.61           & 2.63          \\
\textbf{Q4} & 5                & 907            & 4              & 565           & 1.25           & 1.61          \\
\textbf{Q5} & 83               & 17329          & 67             & 7298          & 1.24           & 2.37          \\
\textbf{Q6} & 69               & 12589          & 58             & 4336          & 1.19           & 2.90          \\
\textbf{Q7} & 50               & 7340           & 38             & 4120          & 1.32           & 1.78          \\         
\bottomrule
\end{tabular}
\end{table}

Depending on the number of intermediate results (reflected by the number of misses without cache), the performance of (\ref{eq:card}) is improved by a factor of 2.61 in the best case. A general trend emerges from the "Ratio" column is that the ratios increase when there are more intermediate results (except Q2). This is inline with the intuition that the more intermediate results there are the higher probability that they overlap. In this evaluation queries of each template are generated without a particular tendency to overlap, while in real-world scenarios the occurrences of queries tend follow more skewed distributions. The LFU cache has the ability to store cardinalities of those mostly used \rss and thus the hit-miss ratio and the performance improvement ratios are likely to be higher than in this evaluation. Further justification of this speculation is planned in future work. 

\section{Accuracy Evaluation}\label{accr}

We use the same set of queries described in Section~\ref{evalquery} to evaluate the accuracy of \pr. Correlation between the estimates given by \pr and the real cardinalities is calculated to measure the accuracy of \pr. We use correlation and not a comparative approach for the following reasons:

\begin{itemize}
\item Correlation measures how much variance of the real cardinalities are explained by the estimates and also works when cardinalities are 0's (as opposed to some relative accuracy measurements~\cite{Neumann2011,HuangHaiLiu}). There are more sophisticated methods for assessing estimation and predication models~\cite{Steyerberg2010} which work best with binary output (true or false) predictions and are not applicable to cardinality estimation.
\item Many existing approaches estimate query components (e.g., individual triple patterns or star patterns) and produce query cardinalities based on predicate independence assumptions. Evaluations~\cite{Neumann2011,HuangHaiLiu} show that these approaches are not very accurate on even simple queries. 
\item Characteristic set is shown to be more accurate~\cite{Neumann2011} among existing approaches but 1) it only addresses star-shape patterns and 2) it utilises a heuristic that only takes the least selective bound node (if any) into account, i.e., only 1 bound node is considered in the estimation even there are more in the query. Most of its evaluation queries have less than 2 bound nodes which \pr gives true cardinalities at runtime within 200 ms.
\end{itemize}

\subsection{Results and Analysis}
For each query template we present the average true cardinalities, average estimates and the correlation between true cardinalities and estimates in Table~\ref{tab:accuracy}.
\begin{table}[!htb]
\centering
\caption{Correlation between true cardinalities and \pr estimates}
\label{tab:accuracy}
\begin{tabular}{@{}rrrr@{}}
            & \multicolumn{2}{c}{Mean} & \\
            \cmidrule(lr){2-3}
            & \textbf{True Card.} & \textbf{\pr} & \textbf{Correl.} \\ \toprule
\textbf{Q1} & 77.5                & 77              & 0.99             \\
\textbf{Q2} & 2.5                 & 0               & 0.05             \\
\textbf{Q3} & 45.3                & 46              & 0.95             \\
\textbf{Q4} & 1                   & 1               & 0.77             \\
\textbf{Q5} & 7.2                 & 2.6             & 0.87             \\
\textbf{Q6} & 6                   & 6               & 1             \\
\textbf{Q7} & 1                   & 1               & 0.87             \\ \bottomrule
\end{tabular}
\end{table}

For most queries the estimates correlate well with the true cardinalities (3 queries $\geq 0.95$, 5 queries $\geq 0.85$, 6 queries $\geq 0.75$), and their averages are close. 

Q1--4 are based on the same templates of the YAGO queries used in characteristic set~\cite{Neumann2011} but have more bound nodes since \pr gives true cardinalities for queries with less than 2 bound nodes. \pr gives good estimates for all but Query~2. Query~2 shows a low correlation for 2 main reasons: 1) there are many 0-valued true cardinalities where \pr gives small but non-zero estimates (most are less than $10^{-9}$ while a few are between 0.01 and 1), and 2) \pr assumes independence of bound nodes but Query~2 does not satisfy this assumption. In contrast Query~4,7 have lower average true cardinalities but \pr gives good estimates. Q6 is a special case where all 3 bound nodes are independent and thus \pr gives perfect predictions.

\subsection{Multipartite Graph and Correlated Nodes}

We use a multipartite-graph model to demonstrate how correlated nodes affect cardinality estimation and how \pr (partially) addresses it.

\begin{figure}[t!]
    \centering
    \begin{subfigure}[t]{0.48\columnwidth}
        \centering
        \includegraphics[width=0.8\columnwidth]{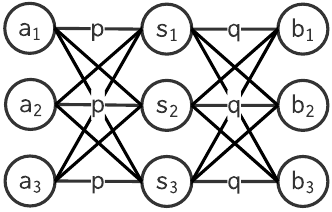}
        \caption{A tripartite graph matched by predicates that form Cartesian products.}\label{fig:subcomplete}
    \end{subfigure}
    ~ 
    \begin{subfigure}[t]{0.48\columnwidth}
        \centering
        \includegraphics[width=0.8\columnwidth]{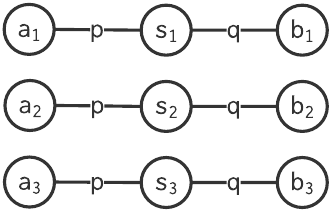}
        \caption{A linear graph matched by two bijective predicates.}\label{fig:sublinear}
    \end{subfigure}%
\end{figure}

Considering a simple query $Q=\{?s~p~?a;~q~?b\}$, if both $p$ and $q$ map each value of $?s$ to all values of $?a$ and $?b$ respectively, then the RDF graph matched by this query is a tripartite graph as illustrated in Figure~\ref{fig:subcomplete}; if $p$ and $q$ are bijective then the matched RDF graph is a linear graph as illustrated in Figure~\ref{fig:sublinear}. They represent two extreme cases in cardinality estimation.

In the former case query results are the Cartesian product of all values of $?a$, $?s$ and $?b$, i.e., $|Q|=|?a|\cdot|?s|\cdot|?b|$. If $?a$ and $?b$ are bound to $a_i$ and $b_j$ respectively, the new cardinality would be $|Q|\cdot sel(?a=a_i)\cdot sel(?b=b_j)$ where $sel(cond)$ gives the selectivity of a condition. Selectivities of $?a$ and $?b$ are independent since the joint selectivity is the product of the two individual selectivities, i.e., $sel(?a=a_i \wedge ?b=b_j)=sel(?a=a_i) \cdot sel(?b=b_j)$. Approaches that calculates joint selectivities similarly assumes an RDF graph as shown in Figure~\ref{fig:subcomplete}.

In the latter case each value uniquely identifies the other two and the query cardinality is given by the equation $|Q|=|?a|\cdot\frac{|?s|}{c}\cdot\frac{|?b|}{c}$ where $c=3$ is the number of (isomorphic) components in Figure~\ref{fig:sublinear}. With bindings $?a=a_i$ and $?b=b_j$ changes the cardinality to $|Q|\cdot sel(?a=a_i)\cdot sel(?b=b_j)\cdot c$ if $a_i$ and $b_j$ are in the same component, 0 otherwise. The joint selectivity is no longer the product of individual selectivities. Characteristic set belongs to this category and it assumes that all bound values are in the same component. We study the structure of YAGO data and find that about 98\% of predicates link a subject to no more than 2 objects. In other word the YAGO data match well with Characteristic set's assumption on short property paths (including star-shapes).

Long property paths are likely to contain predicates that are mixes of the above two cases and heuristics used in many existing approaches are likely to fail. \pr get rid of this issue by calculating the actual number of paths an RDF node and does not assume that all bound nodes happen to be in the same component.

\section{Conclusions}\label{conl}

In this paper we describe \pr that can accurately estimate cardinalities of arbitrary queries with a probabilistic model. \pr avoids issues in existing approaches such as relying on unrealistic assumptions of nodes and predicates correlation or accumulating errors when dealing with complex queries. It utilises memoisation and LFU cache to reduce running time and bound storage usage to a constant limit. Evaluation on YAGO shows that \pr gives estimates that correlate well with true cardinalities.

In the future it is straightforward to compress cardinality caching by utilising a bloom filter, i.e., merging graphs with similar cardinalities in a bucket and using a bloom filter to retrieve the cardinality of a graph. Especially this technology works well when a large proportion of an RDF graph are linear graphs.

\pr does not fully address correlation among values of bound nodes, and in a longer term we aim to address this issue with probabilistic graph models. Also utilising ontology when available would help build a finer probabilistic model for cardinality estimation.

\bibliographystyle{abbrv}
\bibliography{mybib} 

\balance


\end{document}